# Few-shot Class-incremental Audio Classification Using Dynamically Expanded Classifier with Self-attention Modified Prototypes

Yanxiong Li, *Senior Member, IEEE*, Wenchang Cao, Wei Xie, Jialong Li, and Emmanouil Benetos, *Senior Member, IEEE*

*Abstract*—Most existing methods for audio classification assume that the vocabulary of audio classes to be classified is fixed. When novel (unseen) audio classes appear, audio classification systems need to be retrained with abundant labeled samples of all audio classes for recognizing base (initial) and novel audio classes. If novel audio classes continue to appear, the existing methods for audio classification will be inefficient and even infeasible. In this work, we propose a method for few-shot class-incremental audio classification, which can continually recognize novel audio classes without forgetting old ones. The framework of our method mainly consists of two parts: an embedding extractor and a classifier, and their constructions are decoupled. The embedding extractor is the backbone of a ResNet based network, which is frozen after construction by a training strategy using only samples of base audio classes. However, the classifier consisting of prototypes is expanded by a prototype adaptation network with few samples of novel audio classes in incremental sessions. Labeled support samples and unlabeled query samples are used to train the prototype adaptation network and update the classifier, since they are informative for audio classification. Three audio datasets, named NSynth-100, FSC-89 and LS-100 are built by choosing samples from audio corpora of NSynth, FSD-MIX-CLIP and LibriSpeech, respectively. Results show that our method exceeds baseline methods in average accuracy and performance dropping rate. In addition, it is competitive compared to baseline methods in computational complexity and memory requirement. The code for our method is given at https://github.com/vinceasvp/FCAC.

*Index Terms*—Few-shot learning, incremental learning, audio classification, self-attention mechanism, modified prototype

## I. INTRODUCTION

AUDIO classification is a task to recognize different sounds in the environment. Audio classification has been an active research topic with a wide range of applications, including content analysis and retrieval of multimedia [1], [2], audio captioning [3], traffic surveillance [4], bio-acoustic monitoring [5], [6], automatic assisted driving [7], and smart home [8].

This work was supported by the national natural science foundation of China (62111530145, 61771200), international scientific research collaboration project of Guangdong (2021A0505030003), Guangdong basic and applied basic research foundation (2021A1515011454), and a Royal Society International Exchanges grant (ref: IEC\NSFC\201382).

Yanxiong Li, Wenchang Cao, Wei Xie and Jialong Li are with the School of Electronic and Information Engineering, South China University of Technology, Guangzhou, China. (e-mail: eeyxli@scut.edu.cn; wenchangcao98@163.com; chester.w.xie@gmail.com; lijialongjy@163.com). The corresponding author is Dr. Yanxiong Li.

Emmanouil Benetos is with the School of Electronic Engineering and Computer Science, Queen Mary University of London, E1 4NS London, UK (e-mail: emmanouil.benetos@qmul.ac.uk).

### A. Related Works

Many efforts were made on audio classification with the focus on designing discriminative features (e.g., embeddings) or training effective classifiers (e.g., deep neural networks) [9]-[13]. The assumption of most works on audio classification is that the number and type of audio classes to be classified are known in advance. That is, the vocabulary of audio classes is pregiven and fixed. Although these methods are satisfactory in accuracy, they still have shortcomings. For example, the trained classification systems can only recognize the audio classes that are contained in the predefined vocabulary. To recognize novel audio classes, the classification systems have to be retrained using lots of labeled samples of the base and novel audio classes. Retraining the classification systems is laborious and time-consuming for end-users. If the samples of base audio classes are no long available, finetuning the classification system with samples of novel audio classes will make the classification system quickly forget the knowledge of base audio classes, i.e., the problem of catastrophic forgetting [14]. However, the vocabulary of audio classes dynamically changes or expands in many application scenarios, since end-users need to customize the vocabulary according to their preferences. For instance, the end-users often add novel audio classes to audio classification system, such as abnormal sound events, rare musical instruments, audio wake-up words, or animal sounds.

To reduce the demand for the amount of training samples during the construction of classification system, some works were done on few-shot audio classification [15]-[18] and sound event detection [19], [20]. In these methods, the classification system can recognize novel audio classes from a few training samples. Metric-based methods [21] and optimization-based methods [22] are two main lines for few-shot learning. It has been shown that the metric-based method with prototypical networks [21] obtains better results for audio [17]-[19]. Besides the methods based on few-shot learning, Xie et al. [23] investigated zero-shot learning for audio classification using semantic embeddings that are learned from textual labels and sentence descriptions of audio classes. They aimed to recognize the audio classes with only semantic side information and without training samples. Although these methods above can recognize novel audio classes with only few samples, they do not maintain the audio class vocabulary of training samples. As a result, these methods cannot remember the knowledge of old audio classes when they recognize novel ones.

To continually recognize novel classes without forgetting base classes, some researchers proposed the methods based on incremental learning (continual learning, lifelong learning) [24], [25]. There are two main streams in recent works, namely the multi-class incremental learning [25]-[27] and the multi-task incremental learning [28], [29]. The incremental learning has

2been applied in sound event detection [30] and classification [31]-[34] to recognize new sound events without forgetting old ones. Although the methods based on incremental learning can recognize both novel classes and base classes, they still have drawbacks. For example, they typically require to retrain (or update) the classification system with large amounts of labeled samples of novel classes for recognizing novel classes. The requirement for a large number of training samples of novel audio classes is obviously impractical and even infeasible for end-users when the training samples of novel audio classes are few or the computing resources are limited.

As a newly-emerging learning paradigm inspired by cognitive learning, Few-Shot Class-Incremental Learning (FSCIL) is recently proposed [35], [36]. The FSCIL-based methods aim to dynamically expand the capability of the classification system with few training samples in incremental sessions. Although they can combine the strongpoints of the methods of few-shot learning and incremental learning, they confront two challenges that are beyond previous learning paradigms. First, finetuning the classification system with training samples of novel classes leads to catastrophically forgetting the knowledge of base classes. Second, updating with few training samples of novel classes makes the classification system overfit the novel classes. To tackle these two problems above, a decoupled learning scheme is proposed, where a well-trained initial system consists of a feature extractor and a classifier [35]-[41]. For example, Tao et al. [35] designed a neural gas network to preserve the feature topology in the base and novel classes. Gidaris et al. [37] proposed a method of Dynamical Few-Shot Learning (DFSL) by an attention-based weight generator and a cosine-similarity based classifier. Mazumder et al. [38] reduced the complexity of network and alleviated the overfitting problem by squeezing parameters of the neural network. The techniques of Continual Evolution of Classifier (CEC) [39] and knowledge distillation [40] make the classification system memorize base classes and generalize to novel classes. Yang et al. [41] designed a dynamic support network to regularize feature space for overcoming the problems of forgetting and overfitting. Wang et al. applied the DFSL [42] to audio classification. Xie et al. proposed an audio classification method via Adaptively Refined Prototypes (ARP), where a random episodic training strategy and a dynamic relation projection module are used to produce prototypes [43].

Although these efforts above promote the development of the FSCIL technique and benefit to Few-shot Class-incremental Audio Classification (FCAC), we argue that three critical aspects for further performance improvement are largely ignored. First, the unlabeled query samples are not explicitly considered in training and updating the classification system in incremental sessions. Like the labeled support samples, the unlabeled query samples are also informative for updating the classification system. Second, existing works have not paid enough attention to the training of the initial classification system. Current training strategy needs to be optimized for improving the generalization ability of the initial classification system. Third, the knowledge of classifier prototypes in prior incremental sessions is not effectively utilized to update the classifier prototypes in current incremental session.

*B. Our Contributions*

Based on the descriptions above, it is known that many works concerning the FSCIL have been done in the field of computer vision and there are still areas for improvement in these existing works (e.g., the three aspects given in the last paragraph of Section I.A). In addition, the FCAC work has not been carried out so far, which motivates us to address the FCAC problem in this paper. The FCAC task here and the FSCIL task in computer vision have some similarities. For example, they aim to obtain a classification system that can continuously recognize new classes without forgetting the old ones. However, there are differences between them. For instance, the implementation details of these two tasks are different. Specifically, the input features, training strategies and architectures of the classification systems, and performance metrics used in these two tasks are different.

We propose a method for FCAC, which can recognize novel audio classes using few training samples per novel audio class in incremental sessions without forgetting the knowledge of old ones. In the proposed method, we utilize a decoupled training scheme to construct an Embedding Extractor (EE) and then to train a classifier in the base session. We design a Prototype Adaptation Network (PAN) which is used for classifier update and is trained using the samples of base audio classes in the episodic way (i.e., the few-shot learning paradigm). The EE is frozen after construction in base session, whereas the classifier is continually expanded and updated by the PAN in incremental sessions. Because the information learned from both labeled support samples and unlabeled query samples is beneficial for audio classification, both of them, rather than only labeled support samples, are used to train PAN and update classifier. In addition, the classifier prototypes in prior incremental sessions are adopted for updating the counterparts in current incremental session. As a result, the distances among the updated prototypes can be enlarged, and thus the confusions among different audio classes are expected to be reduced.

Three audio datasets, named NSynth-100, FSC-89 and LS-100 are generated by selecting samples from three public audio corpora of the NSynth, FSD-MIX-CLIP and LibriSpeech, respectively. To reproduce our experiments, the generation details (including metadata, explanations) of these three audio datasets are described at https://github.com/vinceasvp/FCAC. Results indicate that our method outperforms baseline methods in terms of Average Accuracy (AA) and Performance Dropping rate (PD), and has advantages over most baseline methods in terms of memory requirement and computational complexity.

In short, the main contributions of the work in this paper are summarized as follows:

1. To continually recognize novel audio classes and remember old audio classes in each incremental session, we design a dynamically expanded classifier with self-attention modified prototypes. The classifier is updated by the PAN. The PAN is a self-attention neural network which can effectively make use of prototypes of prior sessions and unlabeled query samples of current session to update all prototypes of the classifier. Although the basic module of the PAN is similar to the self-attention module used in prior works, on the whole, it is a new network with novel architecture and is specifically designed for updating the classifier. In addition, the update of the classifier using both prior prototypes and unlabeled query samples are not considered in prior works.
2. We propose a Strategy of Training Data Usage (STDU) in base session for training the EE and PAN. The STDU is

specifically designed for our FCAC method, which is also not used in previous works.
3. We propose a pseudo incremental learning strategy for training the PAN in episodic way. The proposed strategy can effectively use the large-scale training dataset of base session to train the PAN which is expected to have strong generalization capability in incremental sessions.
4. We propose a method for solving the problem of FCAC. We comprehensively evaluate the effectiveness of the proposed method, and compare it with baseline methods on three audio datasets from different aspects. Experimental results show that the proposed method has advantages over the baseline methods in terms of both AA and PD.

In summary, the four contributions above make the proposed FCAC method different from the existing FSCIL methods in computer vision, even if these two kinds of methods basically have the same framework (i.e., front-end feature extractor plus back-end classifier). The rest of this paper is organized as follows. Section II introduces the details of the proposed method. Section III presents the experiments and discussions, and the conclusions are finally drawn in Section IV.

## II. METHOD

This section introduces our method, including descriptions of problem definition, whole framework, EE, PAN, and classifier.

### A. Problem Definition

In this work, we focus on the problem of FCAC, which aims to continually recognize novel audio classes from few training samples of novel audio classes without forgetting old ones. The incremental sessions of FCAC come in sequence. Once the update of the classification system enters next session, the training samples in all prior sessions are no longer available. However, the evaluation of the classification system in each session involves audio classes in current and all prior sessions.

The training and evaluation (testing) datasets of different sessions are denoted as $\{\boldsymbol{D}_0^t, \boldsymbol{D}_1^t, \ldots, \boldsymbol{D}_i^t, \ldots, \boldsymbol{D}_{I-1}^t\}$ and $\{\boldsymbol{D}_0^e, \boldsymbol{D}_1^e, \ldots, \boldsymbol{D}_i^e, \ldots, \boldsymbol{D}_{I-1}^e\}$, respectively, where $t$, $e$ and $I$ stand for training, evaluation and total number of sessions, respectively. $\boldsymbol{D}_i^t$ and $\boldsymbol{D}_i^e$ have the same label set which is denoted by $\boldsymbol{L}_i$. The datasets in different sessions do not have the same kind of audio classes, i.e., $\forall i, j$ and $i \neq j, \boldsymbol{L}_i \cap \boldsymbol{L}_j = \emptyset$. In the $i$th session, only $\boldsymbol{D}_i^t$ can be used to train the classification system, and the trained classification system is evaluated on the evaluation datasets of both current and all prior sessions, namely $\boldsymbol{D}_0^e \cup \boldsymbol{D}_1^e \ldots \cup \boldsymbol{D}_i^e$. Session 0 is called base (initial) session, in which the audio classes, dataset $\boldsymbol{D}_0^t$ and classifier are called base audio classes, base training dataset and base classifier, respectively. The dataset $\boldsymbol{D}_0^t$ is a relatively large-scale dataset in which abundant samples per audio class are available to train the classification system. In converse, the datasets $\boldsymbol{D}_1^t$ to $\boldsymbol{D}_{I-1}^t$ in incremental sessions are small-scale datasets, and each of them consists of $N$ audio classes, and each audio class has $K$ training samples. That is, the training dataset $\boldsymbol{D}_i^t$ in the $i$th ($1 \leq i \leq (I-1)$) incremental session is constructed as a $N$-way $K$-shot training dataset. For example, in the FSC-89 dataset, there are 59 audio classes in the base session and each audio class has 800 training samples, whereas 5 audio classes and 5 training samples per audio class are generally available in each incremental session.

### B. Whole Framework

As shown in Fig. 1, the proposed framework includes two kinds of sessions: base and incremental sessions. There are four sequential steps in the base session, namely pre-training EE, training PAN, training EE, and constructing base classifier. In each incremental session, the prototypes of the classifier are updated to recognize novel and old audio classes.

To prevent the audio classification system from forgetting the knowledge of old audio classes in incremental sessions, we decouple the learning of embeddings and the training of classifier. First, we divide $\boldsymbol{D}_0^t$ into two parts with different kinds of audio classes: $\boldsymbol{D}_{0,1}^t$ and $\boldsymbol{D}_{0,2}^t$ (see Tables II to IV for their proportions in $\boldsymbol{D}_0^t$ of three audio datasets). Then, we pre-train an EE in a typically supervised way with $\boldsymbol{D}_{0,1}^t$ where adequate samples per audio class are available. The pre-trained EE can learn discriminative embeddings from Log Mel-spectra of samples for training the PAN. Then, $\boldsymbol{D}_{0,1}^t$ and $\boldsymbol{D}_{0,2}^t$ are used as pseudo base audio class and pseudo novel audio class, respectively, and independently split into many batches. Each batch consists of one support set and one query set. The PAN is episodically trained on each batch. After training on all batches, the PAN is frozen and used to update prototypes of classifier in incremental sessions. Afterwards, the EE is trained on $\boldsymbol{D}_0^t$ in typically supervised way. After training, it is adopted to learn embeddings from Log Mel-spectra of audio samples. Finally, the mean vectors of embeddings of the same kind of audio class are computed and adopted as the prototypes of base classifier.

In base session, $\boldsymbol{D}_0^t$ is split into $\boldsymbol{D}_{0,1}^t$ and $\boldsymbol{D}_{0,2}^t$. Label sets of $\boldsymbol{D}_{0,1}^t$ and $\boldsymbol{D}_{0,2}^t$ are denoted as $\boldsymbol{L}_{0,1}$ and $\boldsymbol{L}_{0,2}$, respectively, and $\boldsymbol{L}_{0,1} \cap \boldsymbol{L}_{0,2} = \emptyset$, $\boldsymbol{L}_{0,1} \cup \boldsymbol{L}_{0,2} = \boldsymbol{L}_0$. The motivation for splitting $\boldsymbol{D}_0^t$ into two parts is based on three considerations. First, to make the PAN have strong generalization ability in incremental sessions, adequate samples of novel audio classes are needed to train the PAN under the incremental learning scenario, namely in the way of $N$-way $K$-shot learning. However, training data of only one session can be used in each incremental session where the amount of training data is quite limited. Second, abundant data is also required to train the EE for making it have strong stability and generalization ability in incremental sessions. Third, in our experiment, it is found that poor results for audio classification are obtained if the EE and PAN are trained with the same kinds of audio classes in $\boldsymbol{D}_0^t$. The reason is probably that the EE pre-trained on the audio classes in $\boldsymbol{D}_0^t$ can represent the audio classes very well and the PAN training on the same kind of audio classes will not acquire useful information for improving the generalization ability of the PAN. The above splitting of $\boldsymbol{D}_0^t$ for pre-training EE and training PAN in base session is called the STDU. The reason why the STDU is designed as described above is that abundant samples in $\boldsymbol{D}_0^t$ can be effectively used to train the EE and PAN, and make them have strong generalization ability in the incremental sessions.

After training the EE, PAN and base classifier in the base session, the framework can be used for incremental learning. In each incremental session, support and query embeddings are learned by the EE from Log Mel-spectra of samples in support and query sets, respectively. Next, embeddings of current session and prototypes of last session are all fed to the PAN for classifier update. Finally, all updated prototypes are used as the dynamically expanded classifier for evaluation.

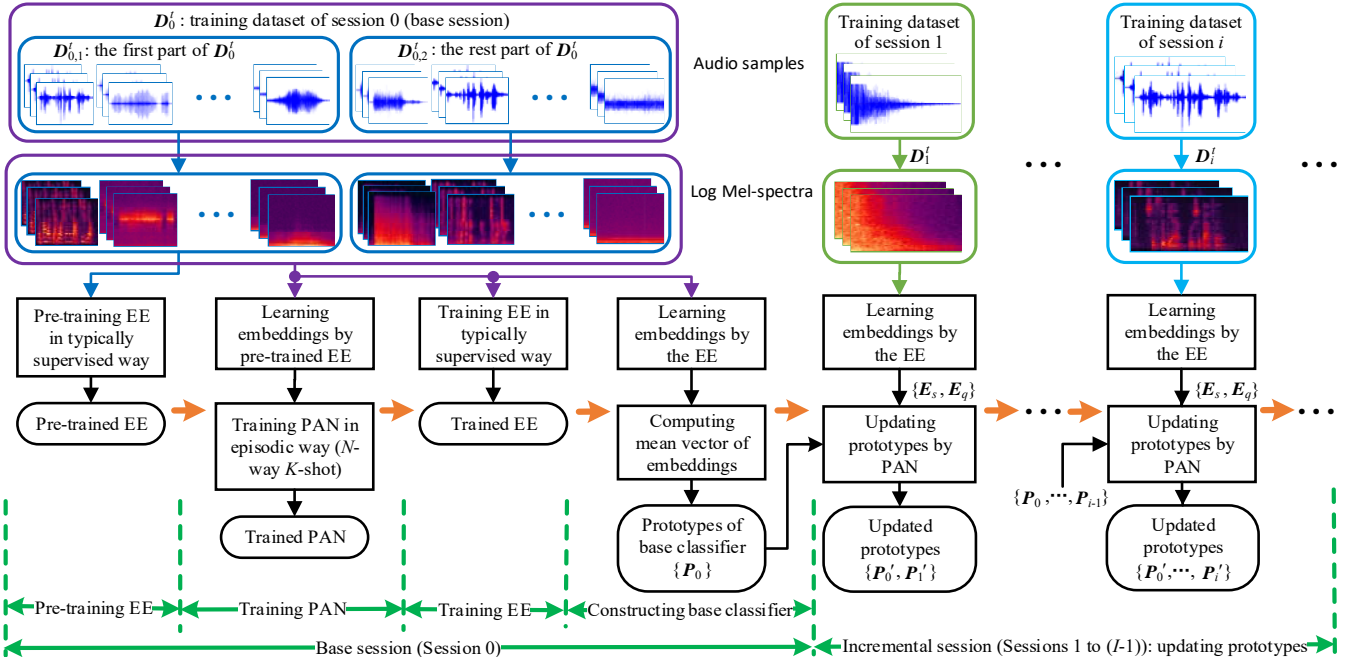

Fig. 1. The proposed framework for FCAC includes two kinds of sessions: base session and incremental session. The embedding extractor, prototype adaptation network, and base classifier are sequentially trained in base session, while prototypes of classifier are updated in each incremental session. EE: embedding extractor; PAN: prototype adaptation network; $P$: prototypes; $E_q$: query embeddings; $E_s$: support embeddings.

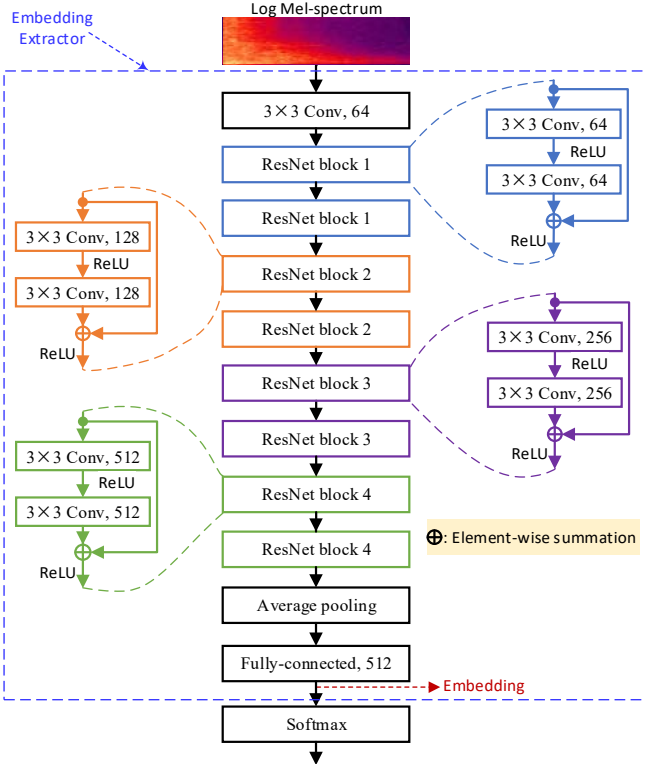

Fig. 2. The architecture of the embedding extractor.

### C. Embedding Extractor

The ResNet [44] based neural network has strong capability to learn discriminative embeddings, and has been successfully applied to many tasks related to audio and video processing [45]-[47]. Inspired by its success for embedding learning in these tasks, the EE adopted in this work is the backbone of a convolutional neural network based on the ResNet.

The architecture of the EE is shown in Fig. 2, which consists of one convolutional layer, eight ResNet blocks with four types of parameters, one average pooling layer, and one Fully Connected (FC) layer. The Softmax layer in the network of Fig. 2 is utilized for training the EE, and is then removed after finishing the EE training. Each ResNet block consists of two convolutional layers followed by the operations of ReLU (Rectified Linear Unit) and element-wise summation. The parameters of each layer and block are presented in Fig. 2. The EE is trained using the Adam optimizer [48] with cross-entropy loss, namely in typically supervised training way. In evaluation stage, the embeddings learned from Log Mel-spectrum of samples are output from the FC layer of the trained EE.

The input of the EE is Log Mel-spectrum which is widely used as input feature of neural network for embedding learning [49]-[51]. Its extraction procedure is briefly introduced as follows. First, each sample is split into overlapping audio frames with fixed length and the audio frames are windowed by a Hamming window. Next, the fast Fourier transformation is conducted on the windowed audio frames to generate power spectrum which is then smoothed by a set of Mel-scale filters. Finally, the logarithm operation is performed on the output of Mel-scale filters to produce the Log Mel-spectrum.

### D. Prototype Adaptation Network

To make the classifier have discriminative decision boundaries over all audio classes, we design a PAN to update prototypes of the classifier in incremental sessions. A prototype is initialized as the mean vector of all support (or training) embeddings of one audio class, and then updated by the PAN. The PAN is a self-attention network, which mainly consists of two modules, namely Attentive Prototype Generation Module (APGM) and Prototype Query-embedding Adaptation Module (PQAM). The APGM is used to generate prototypes of current session from support embeddings. The PQAM is used to update prototypes and query embeddings, whose input is the concatenation of prototypes of current session, query embeddings of current session, and prototypes of last session. The architecture of the PAN is depicted in Fig. 3.

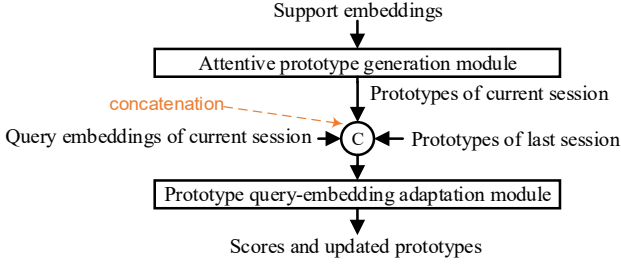

Fig. 3. The architecture of prototype adaptation network, which consists of an attentive prototype generation module (detailed in Fig. 4) and a prototype query-embedding adaptation module (detailed in Fig. 5).

The motivation for designing the PAN as described above is based on the consideration that it is key for the PAN design to effectively use few accessible data to generate representative prototypes of novel classes and adjust the prototypes of all classes in current session. We design the APGM and PQAM mainly using self-attention mechanism [52] since it can acquire the mutual relationship among all input vectors (i.e., support embeddings of novel audio classes for the APGM, prototypes of all audio classes and query embeddings of current session for the PQAM). Hence, the information contained in few accessible samples can be effectively utilized. The generated prototypes of novel audio classes are expected to be representative, and the updated prototypes of all audio classes are hopefully to be separated from each other in the prototype space.

Abundant data of novel audio classes is needed to train the PAN for guaranteeing strong generalization capability of the PAN in incremental sessions. In addition, the base training dataset $D_0^t$ has adequate samples, whereas the incremental training dataset $D_i^t$ has few samples. Hence, we propose a pseudo incremental learning strategy for training the PAN in episodic way on $D_0^t$ to imitate the real evaluation scenario. The proposed strategy is actually a meta-learning-based algorithm [53], where a support set and a query set are a training subset and an evaluation subset, respectively.

The proposed strategy is summarized as Algorithm I, as shown in Table I. We first construct support set $S_b$ and query set $Q_b$ of pseudo base audio classes by randomly choosing samples from $D_{0,1}^t$. Similarly, we construct support set $S_n$ and query set $Q_n$ of pseudo novel audio classes by randomly selecting samples from $D_{0,2}^t$. Both $S_b$ and $S_n$ are merged to form support set $S$, while both $Q_b$ and $Q_n$ are merged to form query set $Q$. Next, embeddings of samples in $S$ and $Q$ are learned using the pre-trained EE. Prototypes of current session are generated by the APGM from support embeddings of $S$. All prototypes and query embeddings are updated using the PQAM whose input is the concatenation of three elements: prototypes of current session, query embeddings of current session, and prototypes of last session. After predicting labels $\hat{L}_q$ for $Q$ by the PQAM based on the updated prototypes and updated query embeddings, cross-entropy loss of $\mathcal{L}(L_q, \hat{L}_q)$ is computed for optimizing the PAN by the algorithm of stochastic gradient descent [54]. The steps above are repeatedly conducted by feeding various batches of support and query sets into the PAN until all samples in $D_0^t$ are selected once. After training, the PAN is frozen and will be utilized to update prototypes of the classifier in real incremental sessions.

TABLE I
ALGORITHM I: PSEUDO INCREMENTAL LEARNING STRATEGY FOR TRAINING THE PAN IN EPISODIC WAY ON $D_0^t$.

**Initialization**:
  $L_q$ and $\hat{L}_q$ are the ground-truth and predicted labels, respectively. $\mathcal{L}(\cdot)$ is the cross-entropy loss. $D_{0,1}^t$ and $D_{0,2}^t$ are the first and rest parts of $D_0^t$, respectively. The EE is pre-trained on $D_{0,1}^t$. The PAN is randomly initialized on $D_0^t$, including the APGM and PQAM.
**While** not done **do**:
  Construct support set $S_b$ and query set $Q_b$ for pseudo base audio classes by randomly selecting samples from $D_{0,1}^t$.
  Construct support set $S_n$ and query set $Q_n$ for pseudo novel audio classes by randomly selecting samples from $D_{0,2}^t$.
  Construct support set $S$ by merging $S_b$ and $S_n$; and construct query set $Q$ by merging $Q_b$ and $Q_n$;
  Learn embeddings for samples in $S$ and $Q$ by the pre-trained EE.
  Generate prototypes of current session by the APGM from support embeddings of $S$.
  Update all prototypes and query-embeddings by the PQAM whose input is the concatenation of prototypes of current session, query-embeddings of current session, and prototypes of last session.
  Predict labels $\hat{L}_q$ for $Q$ by the PQAM based on the updated prototypes and query-embeddings.
  Compute cross-entropy loss of $\mathcal{L}(L_q, \hat{L}_q)$.
  Optimize the PAN using the algorithm of stochastic gradient descent.
**End while**
**Output**:
  A trained PAN, including APGM and PQAM.

*1) Attentive Prototype Generation Module*

The architecture of the APGM is depicted in Fig. 4, whose main part is a self-attention module. The architecture design of the APGM is mainly inspired from the Transformer [52]. The self-attention module in the APGM can acquire the mutual relationship among all support embeddings of novel audio classes. Hence, all generated prototypes of audio classes are expected to be representative, which benefits for generating a classifier with strong discriminative ability.

Support embeddings $E_s$ that are learned from $N_{nov} \times K$ support samples of current session are fed to the APGM, where $N_{nov}$ and $K$ denote number of novel audio classes and number of samples per novel audio class, respectively. Dimension of each embedding is $D$. Three variables of $X_1$, $X_2$ and $X_3$ are obtained by conducting linear transformation on $E_s$, namely $X_1=\Psi_1(E_s)$, $X_2=\Psi_2(E_s)$ and $X_3=\Psi_3(E_s)$. Then, they are further processed by sequentially conducting the operations of matrix multiplication, scale, Softmax, matrix multiplication and linear transformation. That is, the output of the self-attention module in the APGM is $X''$, and is defined by

$$X'' = \Psi_4\left(\text{softmax}\left(\frac{X_1 X_2^T}{\sqrt{D}}\right) X_3\right), \quad (1)$$

where T denotes transpose operation of a matrix; and the scale operation is defined as dividing by a coefficient $\sqrt{D}$, namely $Y' = \frac{Y}{\sqrt{D}} = \frac{X_1 X_2^T}{\sqrt{D}}$. Then, $X''$ and $X$ are element-wisely summed, and their summation is processed by an operation of layer normalization [55] for producing $X'''$. Finally, the prototype of each audio class is generated by computing average of $K$ vectors of $X'''$, and the $K$ vectors belong to the same kind of audio class. After computing the average for all novel audio classes, $N_{nov}$ prototypes of current session, $P_{nov}$, are obtained.

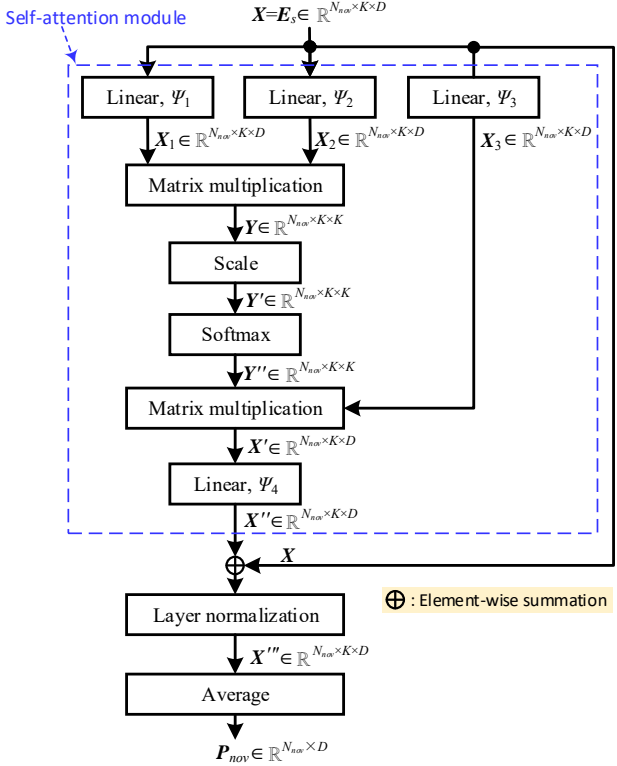

Fig. 4. The architecture of the APGM. $E_s$: support embeddings; $N_{nov}$: number of novel audio classes; $K$: number of support samples per audio class; $D$: dimension of each embedding (prototype); $P_{nov}$: prototypes of novel audio classes.

The parameters of four Linear layers (linear transformations from $\Psi_1$ to $\Psi_4$) are the tunable parameters of the APGM. Their parameters are iteratively tuned during the procedure of PAN training using the pseudo incremental learning strategy.

*2) Prototype Query-embedding Adaptation Module*

With the increase of prototypes of audio class, the prototype space will become crowded, which is unfavorable for audio classification. When audio classes come with groups in the incremental sessions, the prototypes generated in each session may only identify audio classes of current session satisfactorily. When audio classes of all prior sessions and current session are involved in evaluation, direct concatenation of prototypes cannot produce a classifier with strong discrimination ability and thus classification results will be unsatisfactory. To obtain discriminative decision boundaries over all novel and old audio classes, it is important for classifier update to acquire global location information of all prototypes in the prototype space. In addition, query embeddings of unlabeled query samples are also informative and should be used for classifier update. To reach these targets above, we design a PQAM to update prototypes of both old and novel audio classes, and to compute scores (cosine similarities) between query embeddings and prototypes. The architecture of the PQAM is depicted in Fig. 5.

The PQAM is similar to the APGM with minor differences. It also mainly consists of a self-attention module which can acquire the mutual relationship among prototypes of novel audio classes, prototypes of old audio classes and query embeddings. Hence, all updated prototypes of both novel and old audio classes are expected to be far apart in the prototype space, and the updated classifier will provide discriminative decision boundaries over all novel and old audio classes. To efficiently update the prototypes and compute the scores, $K_q$ query embeddings $E_q=\{E_{q,k}\}$, $1 \le k \le K_q$, are simultaneously processed by the PQAM in practice. As illustrated in Fig. 5, the input variable of the PQAM is $X \in \mathbb{R}^{K_q \times (N_{o-n}+1) \times D}$ which is composed of $K_q$ copies of the concatenation of $N_{old}$ prototypes of $P_{old}$, $N_{nov}$ prototypes of $P_{nov}$ and one embedding of $E_{q,k}$. The output of the self-attention module in the PQAM is $X''$ which can be also defined by Eq. (1). Afterwards, $X''$ and $X$ are element-wisely summed, followed by the operation of layer normalization [55] for obtaining $X'''$. Finally, the $X'''$ is split into two parts: updated query embeddings $E'_q$ and updated prototypes $P'$. The cosine similarities (scores) between each updated query embedding $E'_{q,k}$ and $P'$ is used to make decision in evaluation stage for each query sample.

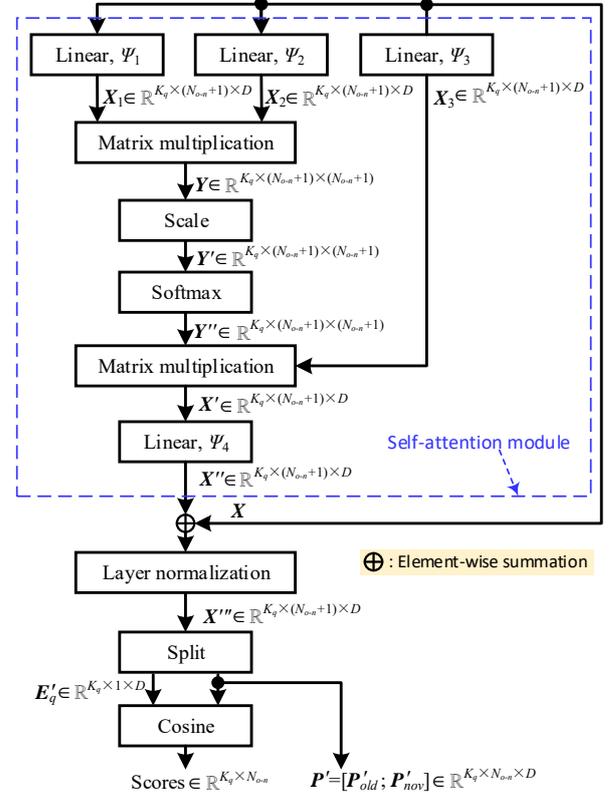

Fig. 5. The architecture of the PQAM. $K_q$: number of query samples; $P_{old}$: prototypes of old audio classes; $P_{nov}$: prototypes of novel audio classes; $E_q$: query embeddings; $E_{q,k}$: embedding of the $k$th query sample; $N_{o-n}$: number of old and novel audio classes; $N_{old}$: number of old audio classes; $N_{nov}$: number of novel audio classes; $D$: dimension of each embedding (prototype).

Together with the parameters of four Linear layers of the APGM, the parameters of four Linear layers of the PQAM are iteratively tuned during the procedure of PAN training using the pseudo incremental learning strategy.

*E. Dynamically Expanded Classifier*

The classifier consists of prototypes and each prototype stands for one audio class (i.e., one prototype per audio class). As depicted in Fig. 1, each prototype of base classifier is obtained by computing the mean vector of embeddings of the same kind of audio class. Each base audio class has abundant training samples, and thus the mean vector of embeddings of one base audio class can represent the audio class well. However, the number of training samples of each novel audio class is few, and thus simply using the mean vector of embeddings of few samples cannot effectively represent the differences between different kinds of audio classes. Hence, the classifier will produce unsatisfactory results in incremental sessions.

Here, we design a dynamically expanded classifier whose prototypes are expanded and updated by the PAN in each incremental session. The updating process of prototypes in each incremental session (real incremental learning) is the same as the training process of the PAN in base session (pseudo incremental learning), except for the samples they use. The samples that are used for updating the prototypes in incremental sessions and for training the PAN in base session are from the $\boldsymbol{D}_i^t$ and the $\boldsymbol{D}_0^t$, respectively. After updating all prototypes, the classifier can provide discriminative decision boundaries over all audio classes, and will perform well for audio classification.

## III. EXPERIMENTS AND DISCUSSIONS

In this section, experimental data and setups are first introduced. Then, we present ablation experiments and comparisons of different methods. Finally, we discuss generalization across audio datasets and conduct extended analyses for our method.

### A. Experimental Datasets

Experiments are performed on the datasets selected from three audio corpora, including FSD-MIX-CLIPS [56], NSynth [57], and LibriSpeech [58]. These three audio corpora are publicly available for research purposes and have been commonly used in previous works for audio classification.

The FSD-MIX-CLIPS is a programmatically mixed audio corpus, where the polyphony and signal-noise-ratio properties are controlled. In this audio corpus, there are 89 sound events (audio classes) and 614 K audio clips (samples). The length of each audio clip is 1 second. The set of audio classes covers a diverse range of real-world sounds, from human and animal sounds to natural, musical or miscellaneous sounds.

The NSynth is a large-scale audio corpus of musical notes. It includes 306,043 musical notes, and each of these notes is with a unique pitch, timbre, and envelope. There are 306,043 audio snippets in the NSynth, and each audio snippet is of four seconds for representing one type of instruments. There are 1,006 instruments (audio classes) in total in this audio corpus.

The LibriSpeech is a speech corpus of approximately 1,000 hours of audiobooks that are spoken by 2,484 speakers (audio classes). Training data is divided into 3 parts of 100 hours, 360 hours, and 500 hours. Development data and testing data are divided into the *clean* and *other* classes, respectively.

Audio datasets that are built from the FSD-MIX-CLIPS, NSynth and LibriSpeech, are denoted as FSC-89, NSynth-100 and LS-100, respectively. They are independently divided into two parts without overlaps of audio classes, namely base dataset $\boldsymbol{D}_0$ and incremental dataset $\boldsymbol{D}_i$ (1≤$i$≤($I$-1)). The base dataset $\boldsymbol{D}_0$ consists of base training dataset $\boldsymbol{D}_0^t$ and base evaluation dataset $\boldsymbol{D}_0^e$, while the incremental dataset $\boldsymbol{D}_i$ is composed of incremental training dataset $\boldsymbol{D}_i^t$ and incremental evaluation dataset $\boldsymbol{D}_i^e$. $\boldsymbol{D}_0^t$ is adopted to train the EE and the PAN in base session, while $\boldsymbol{D}_i^t$ is used to update prototypes of the classifier for evaluation. When $\boldsymbol{D}_0^t$ is used to train the EE in typically supervised way, its samples are fed to the EE as a whole instead of being divided into many small batches. When $\boldsymbol{D}_0^t$ is used to train the PAN in episodic way, its samples are split into two parts: $\boldsymbol{D}_{0,1}^t$ (pseudo base audio classes) and $\boldsymbol{D}_{0,2}^t$ (pseudo novel audio classes). $\boldsymbol{D}_{0,1}^t$ and $\boldsymbol{D}_{0,2}^t$ are independently divided into many batches and each batch is composed of a support set and a query set. When $\boldsymbol{D}_i^t$ is used to update prototypes in episodic way, its samples are split into many batches and each batch consists of a support set and a query set.

In each episodic training stage, $N·K$ samples are randomly chosen from $N$ audio classes ($K$ samples per audio class) in the training dataset ($\boldsymbol{D}_0^t$ or $\boldsymbol{D}_i^t$) to construct the support set, and then $N·K_q$ different samples of the same $N$ audio classes ($K_q$ samples per audio class) are also randomly selected from the training dataset to generate the query set. The selections of both audio classes and samples per audio class are repeated until all audio classes and their samples in the training dataset are chosen once. The selected samples in different batches are different to each other. In evaluation stage, all evaluation datasets are fed to the EE and updated classifier as a whole rather than being divided into many small batches. Tables II, III and IV present the detailed information of the FSC-89, NSynth-100 and LS-100, respectively.

TABLE II
DETAILED INFORMATION OF FSC-89

| Parameters | $\boldsymbol{D}_0$ | | | $\boldsymbol{D}_i$ | |
|---|---|---|---|---|---|
| | $\boldsymbol{D}_{0,1}^t$ | $\boldsymbol{D}_{0,2}^t$ | $\boldsymbol{D}_0^e$ | $\boldsymbol{D}_i^t$ | $\boldsymbol{D}_i^e$ |
| #Classes | 39 | 20 | 59 | 30 | 30 |
| #Samples | 31200 | 16000 | 11800 | 15000 | 6000 |
| Length (hours) | 8.67 | 4.44 | 3.28 | 4.17 | 1.67 |
| #Samples/Class | 800 | 800 | 200 | 500 | 200 |

#Samples/Class: number of samples per audio class; $\boldsymbol{D}_i$: total data of all audio classes in all incremental sessions, 1≤$i$≤($I$-1).

TABLE III
DETAILED INFORMATION OF NSYNTH-100

| Parameters | $\boldsymbol{D}_0$ | | | $\boldsymbol{D}_i$ | |
|---|---|---|---|---|---|
| | $\boldsymbol{D}_{0,1}^t$ | $\boldsymbol{D}_{0,2}^t$ | $\boldsymbol{D}_0^e$ | $\boldsymbol{D}_i^t$ | $\boldsymbol{D}_i^e$ |
| #Classes | 30 | 25 | 55 | 45 | 45 |
| #Samples | 6000 | 5000 | 5500 | 4500 | 4500 |
| Length (hours) | 6.67 | 5.56 | 1.52 | 5.00 | 5.00 |
| #Samples/Class | 200 | 200 | 100 | 100 | 100 |

TABLE IV
DETAILED INFORMATION OF LS-100

| Parameters | $\boldsymbol{D}_0$ | | | $\boldsymbol{D}_i$ | |
|---|---|---|---|---|---|
| | $\boldsymbol{D}_{0,1}^t$ | $\boldsymbol{D}_{0,2}^t$ | $\boldsymbol{D}_0^e$ | $\boldsymbol{D}_i^t$ | $\boldsymbol{D}_i^e$ |
| #Classes | 35 | 25 | 60 | 40 | 40 |
| #Samples | 17500 | 12500 | 6000 | 20000 | 4000 |
| Length (hours) | 9.72 | 6.94 | 3.33 | 11.11 | 2.22 |
| #Samples/Class | 500 | 500 | 100 | 500 | 100 |

### B. Experimental Setup

All experiments are carried out on a machine whose main configurations are as follows: two CPUs of Intel Xeon 8124M with 3.5 GHz, a RAM of 128 GB, and three GPUs of RTX3090. The metric of accuracy is defined as the number of correctly classified samples divided by the total number of samples involved in classification, which is adopted to evaluate the performance of different methods in each session. The metrics of average accuracy AA and performance dropping rate PD are used to measure the overall performance of different methods. They are defined by

$$AA = \frac{1}{I}\sum_{i=0}^{I-1} A_i, \quad (2)$$

$$\begin{cases} PD = A_0 - A_{I-1}, & \text{for the classes of Base and Both} \\ PD = A_1 - A_{I-1}, & \text{for the classes of Novel} \end{cases} \quad (3)$$

where $A_i$ denotes the accuracy in session $i$. The higher the AA or the lower the PD, the better the performance of the methods. Besides, computational complexity and memory requirement of different methods in incremental sessions are measured by the metrics of Average Training Time (ATT) and Storage Space (SS), respectively. The ATT is defined as the average time required to train the classification system in all incremental sessions. The SS is defined as the memory space used to store samples (or embeddings) and parameters of the classification system. The lower the ATT and the SS, the lower computational complexity and the memory requirement of the methods, respectively.

The framework for FCAC mainly includes the EE, PAN and classifier. Its main parameters are given in Table V.

TABLE V
SETTINGS OF MAIN PARAMETERS OF THE PROPOSED FRAMEWORK

| Type | Parameters settings |
|---|---|
| EE | Frame length/overlapping: 25ms/10ms<br>Dimension of log Mel-spectrum: 128<br>Dimension of embedding: 512<br>Learning rate: 0.1 |
| PAN | Number of pseudo base classes: 39 (FSC-89), 30 (NSynth-100), 35 (LS-100)<br>Number of pseudo novel classes: 20 (FSC-89), 25 (NSynth-100), 25 (LS-100)<br>Number of classes in each batch, $N$: 1 to 20<br>Number of support samples per class, $K$: 1 to 20<br>Number of query samples per class, $K_q$: 15<br>Learning rate: 0.0002 |
| Classifier | Dimension of prototypes: 512 |

### C. Ablation Experiments

In this subsection, we conduct ablation analyses to assess the effectiveness of main components of the proposed method. The NSynth-100 is used as experimental dataset in this experiment for simplicity.

As described in section II.B, the datasets for pre-training EE and training PAN are $\boldsymbol{D}_{0,1}^t$ and $\boldsymbol{D}_0^t$, respectively. $\boldsymbol{D}_0^t$ is further divided into $\boldsymbol{D}_{0,1}^t$ and $\boldsymbol{D}_{0,2}^t$ to generate pseudo base classes and pseudo novel classes, respectively. That is, the datasets used for pre-training EE and training PAN are different to each other. The method of data construction for pre-training EE and training PAN is called the STDU. In addition, APGM and PQAM are proposed to generate prototypes of novel audio classes and to update prototypes of all audio classes, respectively. We discuss the impacts of STDU, APGM and PQAM on the performance of the proposed method. In this experiment, the value of ($N$, $K$) is set to (5, 5) without losing generality.

The results obtained by our method on NSynth-100 with different combinations of STDU, APGM and PQAM are listed in Table VI. When all modules of STDU, APGM and PQAM are used in the proposed framework, our method achieves the best performance. The highest AA score of 93.31% and the lowest PD score of 12.90% are obtained for the Both (both Base and Novel). In addition, by comparing the case of ① to the cases of ②, ③ and ⑤ in Table VI, it can be known that each one of these three modules above has contribution to the performance improvement of our method in AA and PD.

### D. Comparison of Different Methods

In this subsection, we compare our method with five baseline methods for FCAC in AA and PD. The baseline methods are denoted as Finetune [59], iCaRL [60], DFSL [42], ARP [43], and CEC [39], and are widely used in previous related works.

TABLE VI
RESULTS OBTAINED BY OUR METHOD ON NSYNTH-100 WITH DIFFERENT COMBINATIONS OF STDU, APGM AND PQAM

| No. | STDU | APGM | PQAM | Session | 0 (0-54) | 1 (55-59) | 2 (60-64) | 3 (65-69) | 4 (70-74) | 5 (75-79) | 6 (80-84) | 7 (85-89) | 8 (90-94) | 9 (95-99) | AA (%) | PD (%) |
|---|---|---|---|---|---|---|---|---|---|---|---|---|---|---|---|---|
| ① | × | × | × | Base | 99.75 | 99.54 | 99.28 | 98.24 | 98.02 | 98.14 | 98.07 | 97.83 | 97.79 | 97.74 | 98.44 | 2.01 |
|   |   |   |   | Novel | - | 71.26 | 71.96 | 73.58 | 69.95 | 69.04 | 65.95 | 62.85 | 63.63 | 62.44 | 67.85 | 8.82 |
|   |   |   |   | Both | 99.75 | 97.18 | 95.07 | 92.96 | 90.54 | 89.04 | 86.73 | 84.22 | 83.41 | 81.86 | 90.08 | 17.89 |
| ② | × | × | √ | Base | 99.93 | 99.77 | 99.79 | 98.91 | 98.38 | 98.66 | 98.54 | 98.36 | 98.33 | 98.31 | 98.90 | 1.62 |
|   |   |   |   | Novel | - | 75.84 | 76.65 | 76.73 | 72.09 | 69.75 | 67.26 | 65.81 | 66.53 | 65.13 | 70.64 | 10.71 |
|   |   |   |   | Both | 99.93 | 97.78 | 96.23 | 94.16 | 91.37 | 89.63 | 87.50 | 85.70 | 84.84 | 83.38 | 91.06 | 16.55 |
| ③ | × | √ | × | Base | 99.96 | 99.87 | 99.91 | **99.29** | **99.24** | **99.30** | **99.26** | 99.24 | 99.21 | **99.23** | **99.45** | **0.73** |
|   |   |   |   | Novel | - | 71.06 | 71.61 | 72.37 | 69.17 | 69.20 | 66.93 | 64.78 | 65.28 | 63.58 | 68.22 | 7.48 |
|   |   |   |   | Both | 99.96 | 97.47 | 95.56 | 93.53 | 91.22 | 89.90 | 87.85 | 85.84 | 84.92 | 83.19 | 90.94 | 16.77 |
| ④ | × | √ | √ | Base | 99.87 | 99.56 | 99.44 | 98.68 | 98.44 | 98.00 | 97.94 | 97.96 | 97.92 | 97.87 | 98.57 | 2.00 |
|   |   |   |   | Novel | - | 78.22 | 84.60 | **82.22** | 80.11 | 77.59 | 73.28 | 71.44 | 71.04 | 68.43 | 76.33 | 9.79 |
|   |   |   |   | Both | 99.87 | 97.78 | 97.16 | 95.15 | 93.55 | 91.62 | 89.23 | 87.65 | 86.60 | 84.62 | 92.32 | 15.25 |
| ⑤ | √ | × | × | Base | 99.91 | 99.78 | 99.80 | 99.10 | 99.06 | 98.96 | 98.82 | 98.88 | 98.82 | 98.84 | 99.20 | 1.07 |
|   |   |   |   | Novel | - | 70.32 | 76.53 | 75.43 | 74.62 | 72.87 | 68.00 | 66.14 | 65.94 | 64.09 | 70.44 | 6.23 |
|   |   |   |   | Both | 99.91 | 97.32 | 96.22 | 94.03 | 92.54 | 90.81 | 87.94 | 86.15 | 84.97 | 83.20 | 91.31 | 16.71 |
| ⑥ | √ | × | √ | Base | **99.98** | 99.92 | 99.90 | 99.25 | 99.10 | 99.29 | 99.19 | **99.27** | 99.25 | 99.23 | 99.44 | 0.75 |
|   |   |   |   | Novel | - | 73.04 | 79.38 | 77.91 | 76.43 | 74.86 | 70.68 | 69.27 | 70.48 | 67.91 | 73.33 | **5.13** |
|   |   |   |   | Both | **99.98** | 97.68 | 96.75 | 94.68 | 93.06 | 91.66 | 89.13 | 87.60 | 87.13 | 85.14 | 92.28 | 14.84 |
| ⑦ | √ | √ | × | Base | 99.95 | 99.84 | 99.87 | 99.21 | 98.96 | 98.64 | 98.56 | 98.49 | 98.46 | 98.45 | 99.04 | 1.50 |
|   |   |   |   | Novel | - | 80.18 | 85.56 | 80.52 | 78.18 | 78.67 | 73.49 | 71.51 | 70.94 | 68.53 | 76.40 | 11.65 |
|   |   |   |   | Both | 99.95 | 98.20 | 97.67 | **95.21** | 93.42 | 92.40 | 89.71 | 87.99 | 86.87 | 84.98 | 92.64 | 14.97 |
| ⑧ | √ | √ | √ | Base | **99.98** | **99.95** | **99.92** | 99.26 | 98.88 | 98.78 | 98.72 | 98.75 | 98.74 | 98.71 | 99.17 | 1.27 |
|   |   |   |   | Novel | - | **80.39** | **95.45** | 80.24 | **90.58** | **79.46** | **75.69** | **74.30** | **73.89** | **72.86** | **78.09** | 7.53 |
|   |   |   |   | Both | **99.98** | **98.32** | **97.69** | 95.18 | **94.00** | **92.73** | **90.59** | **89.24** | **88.28** | **87.08** | **93.31** | 12.90 |

×: the module is not used. √: the module is used. 0 (0-54): session 0 whose numbers of audio classes are from 0 to 54. Both: base and novel audio classes.

The baseline methods are briefly introduced as follows. In the Finetune method, the classification system can quickly adapt to novel audio classes after finetuning using training samples of novel audio classes in incremental sessions. As a result, the classification system tends to overfit the novel audio classes and forget the old ones. In the iCaRL method, both strong classifiers and a feature representation are learned using the strategies of data retention and knowledge distillation. When novel audio classes appear continually, the classification system tends to gradually forget the old audio classes. In the DFSL method, an attention-based weight generator and a cosine-similarity based classifier are designed for realizing FCAC. In the ARP method, the prototypes are adaptively refined by a dynamic relation projection module. In the CEC method, continually evolved classifiers are designed for recognizing novel audio classes and a graph model is used to propagate the context information between classifiers for prototype adaptation. Based on the introductions above, main technical differences of different methods are presented in Table VII.

TABLE VII
SUMMARY OF DIFFERENT METHODS FOR FCAC

| Methods | Main merits |
|---|---|
| Finetune | Tune existing system with samples of novel audio classes |
| iCaRL | Data retention and knowledge distillation |
| DFSL | Attention-based weight generator; cosine-similarity based classifier |
| ARP | Adaptively refine prototypes by a dynamic relation projection module |
| CEC | Continually evolved classifier; a graph model for prototype adaptation |
| Ours | Classifier with self-attention modified prototypes; data usage strategy |

All baseline methods are implemented with open-source codes by the authors, whose main parameters are set according to the suggestions in the corresponding references and optimally tuned on the training data. Different methods are compared on three audio datasets. In this experiment, the value of ($N$, $K$) is set to (5, 5). Under the same experimental conditions, the scores of accuracies, AA and PD that are obtained by different methods on the audio datasets of FSC-89, NSynth-100 and LS-100 are presented in Tables VIII, IX and X, respectively.

TABLE VIII
RESULTS OBTAINED BY DIFFERENT METHODS ON FSC-89

| Methods | Session | 0 (0-58) | 1 (59-63) | 2 (64-68) | 3 (69-73) | 4 (74-78) | 5 (79-83) | 6 (84-88) | AA (%) | PD (%) |
|---|---|---|---|---|---|---|---|---|---|---|
| Finetune | Base | 42.28 | 31.84 | 29.15 | 24.98 | 20.54 | 19.63 | 16.19 | 26.37 | 26.19 |
| | Novel | - | 29.70 | 28.85 | 23.23 | 25.28 | 21.44 | 17.72 | 24.37 | 11.98 |
| | Both | 42.28 | 31.67 | 29.11 | 24.63 | 21.74 | 20.17 | 16.70 | 26.61 | 25.58 |
| iCaRL | Base | 42.48 | 33.02 | 32.14 | 28.24 | 24.78 | 19.98 | 21.65 | 28.90 | 20.83 |
| | Novel | - | 25.10 | 19.05 | 17.10 | 19.43 | 19.30 | 15.08 | 19.18 | 10.02 |
| | Both | 42.48 | 32.40 | 30.25 | 25.99 | 23.42 | 19.78 | 19.44 | 27.68 | 23.04 |
| DFSL | Base | 42.36 | 36.58 | 36.23 | 35.97 | 35.76 | 35.66 | 35.55 | 36.87 | 6.81 |
| | Novel | - | 19.40 | 12.00 | 12.50 | 12.15 | 11.62 | 11.38 | 13.17 | 8.02 |
| | Both | 42.36 | 35.23 | 32.72 | 31.21 | 29.79 | 28.51 | 27.40 | 32.46 | 14.96 |
| ARP | Base | 42.04 | **41.36** | **39.52** | 38.40 | 37.37 | 36.67 | 36.05 | 38.77 | 5.99 |
| | Novel | - | 23.35 | 22.19 | 20.05 | 19.99 | 19.14 | 18.36 | 20.51 | **4.99** |
| | Both | 42.04 | 39.95 | 37.01 | 34.68 | 32.97 | 31.45 | 30.09 | 35.46 | 11.95 |
| CEC | Base | 42.16 | 40.56 | 39.39 | **38.83** | **38.28** | **37.87** | **37.57** | **39.24** | **4.59** |
| | Novel | - | 31.31 | 22.08 | 23.03 | 25.09 | 25.40 | 23.90 | 25.13 | 7.41 |
| | Both | 42.16 | 39.84 | 36.88 | 35.63 | 34.94 | 34.15 | 32.96 | 36.65 | 9.20 |
| Ours | Base | **42.92** | 40.16 | 38.98 | 38.41 | 37.61 | 37.17 | 36.62 | 38.84 | 6.30 |
| | Novel | - | **38.32** | **31.11** | **32.80** | **34.15** | **32.84** | **30.97** | **33.37** | 7.35 |
| | Both | **42.92** | **40.01** | **37.84** | **37.27** | **36.73** | **35.88** | **34.71** | **37.91** | 8.21 |

TABLE IX
RESULTS OBTAINED BY DIFFERENT METHODS ON NSYNTH-100

| Methods | Session | 0 (0-54) | 1 (55-59) | 2 (60-64) | 3 (65-69) | 4 (70-74) | 5 (75-79) | 6 (80-84) | 7 (85-89) | 8 (90-94) | 9 (95-99) | AA (%) | PD (%) |
|---|---|---|---|---|---|---|---|---|---|---|---|---|---|
| Finetune | Base | 99.96 | 88.91 | 85.41 | 80.36 | 72.51 | 45.24 | 59.31 | 48.53 | 50.68 | 53.28 | 68.42 | 46.68 |
| | Novel | - | 38.75 | 30.25 | 36.96 | 37.54 | 28.95 | 27.24 | 22.30 | 20.58 | 19.00 | 29.06 | 19.75 |
| | Both | 99.96 | 84.73 | 76.92 | 71.06 | 63.18 | 40.15 | 47.99 | 38.33 | 38.01 | 37.86 | 59.82 | 62.1 |
| iCaRL | Base | **99.98** | 98.42 | 99.25 | 98.40 | 94.56 | 82.36 | 85.09 | 80.59 | 75.78 | 74.53 | 88.90 | 25.45 |
| | Novel | - | 36.94 | 31.88 | 35.03 | 38.33 | 35.27 | 30.76 | 26.75 | 25.52 | 22.27 | 31.42 | 14.67 |
| | Both | **99.98** | 93.30 | 88.88 | 84.82 | 79.57 | 67.65 | 65.92 | 59.65 | 54.62 | 51.01 | 74.54 | 48.97 |
| DFSL | Base | 99.93 | 99.11 | 98.83 | 95.83 | 94.84 | 94.81 | 94.39 | 93.76 | 92.06 | 91.61 | 95.52 | 8.32 |
| | Novel | - | 57.01 | 55.57 | 59.89 | 59.35 | 56.46 | 52.29 | 50.94 | 52.57 | 52.49 | 55.17 | **4.52** |
| | Both | 99.93 | 96.00 | 92.95 | 89.26 | 86.47 | 83.66 | 80.28 | 77.68 | 76.12 | 75.01 | 85.74 | 24.92 |
| ARP | Base | 99.96 | 99.85 | 99.72 | 99.10 | 98.58 | 98.56 | 98.50 | 98.37 | 98.23 | 98.15 | 98.90 | 1.81 |
| | Novel | - | 52.99 | 59.92 | 66.25 | 67.62 | 68.41 | 63.54 | 60.05 | 60.35 | 57.17 | 61.81 | -4.13 |
| | Both | 99.96 | 95.95 | 93.60 | 92.06 | 90.32 | 89.14 | 86.16 | 83.47 | 82.28 | 79.69 | 89.26 | 20.27 |
| CEC | Base | 99.96 | 99.87 | 99.90 | **99.29** | **99.24** | **99.30** | **99.26** | **99.24** | **99.20** | **99.23** | **99.45** | **0.73** |
| | Novel | - | 71.06 | 71.61 | 72.37 | 69.17 | 69.20 | 66.92 | 64.80 | 65.28 | 63.59 | 68.22 | 7.47 |
| | Both | 99.96 | 97.47 | 95.56 | 93.52 | 91.22 | 89.90 | 87.85 | 85.84 | 84.92 | 83.19 | 90.94 | 16.77 |
| Ours | Base | **99.98** | **99.95** | **99.92** | 99.26 | 98.88 | 98.78 | 98.72 | 98.75 | 98.74 | 98.71 | 99.17 | 1.27 |
| | Novel | - | **80.39** | **95.45** | **80.24** | **90.58** | **79.46** | **75.69** | **74.30** | **73.89** | **72.86** | **78.09** | 7.53 |
| | Both | **99.98** | **98.32** | **97.69** | **95.18** | **94.00** | **92.73** | **90.59** | **89.24** | **88.28** | **87.08** | **93.31** | 12.90 |

TABLE X
RESULTS OBTAINED BY DIFFERENT METHODS ON LS-100

| Methods | Session | 0 (0-59) | 1 (60-64) | 2 (65-69) | 3 (70-74) | 4 (75-79) | 5 (80-84) | 6 (85-89) | 7 (90-94) | 8 (95-99) | AA (%) | PD (%) |
|---|---|---|---|---|---|---|---|---|---|---|---|---|
| Finetune | Base | 92.02 | 72.90 | 37.03 | 28.12 | 20.75 | 14.45 | 5.70 | 3.23 | 0.27 | 30.50 | 91.75 |
|  | Novel | - | 86.60 | 31.50 | 28.87 | 25.45 | 24.24 | 18.17 | 13.46 | 11.80 | 30.01 | 74.80 |
|  | Both | 92.02 | 73.95 | 36.24 | 28.27 | 21.93 | 17.33 | 9.86 | 7.00 | 4.88 | 32.39 | 87.14 |
| iCaRL | Base | 92.02 | 80.80 | 73.18 | 58.45 | 26.95 | 16.93 | 32.58 | 29.53 | 26.38 | 48.54 | 65.64 |
|  | Novel | - | 58.00 | 67.10 | 57.40 | 20.05 | 16.48 | 30.33 | 26.83 | 28.95 | 38.14 | 29.05 |
|  | Both | 92.02 | 79.05 | 72.31 | 58.24 | 25.23 | 16.80 | 31.83 | 28.54 | 27.41 | 47.94 | 64.61 |
| DFSL | Base | 91.93 | 91.93 | **91.88** | **91.85** | **91.83** | **91.86** | **91.85** | **91.85** | **91.84** | **91.87** | **0.09** |
|  | Novel | - | 53.60 | 61.90 | 50.67 | 48.90 | 51.56 | 47.97 | 44.11 | 45.38 | 50.51 | **8.22** |
|  | Both | 91.93 | 88.97 | 87.60 | 83.61 | 81.11 | 80.01 | 77.22 | 74.26 | 73.25 | 81.99 | 18.68 |
| ARP | Base | **92.35** | **92.22** | 91.85 | 91.66 | 91.60 | 91.56 | 91.47 | 91.43 | 91.31 | 91.72 | 1.04 |
|  | Novel | - | 42.92 | 44.99 | 41.00 | 38.09 | 37.75 | 35.57 | 32.91 | 31.89 | 38.14 | 11.03 |
|  | Both | **92.35** | 88.43 | 85.16 | 81.53 | 78.22 | 75.73 | 72.84 | 69.87 | 67.54 | 79.07 | 24.81 |
| CEC | Base | 91.72 | 91.67 | 91.25 | 91.14 | 91.10 | 91.07 | 90.97 | 90.66 | 90.72 | 91.14 | 1.00 |
|  | Novel | - | 86.30 | 82.76 | 69.67 | 68.25 | 67.06 | 66.03 | 60.35 | 60.05 | 70.06 | 26.25 |
|  | Both | 91.72 | 91.25 | 90.04 | 86.84 | 85.38 | 84.01 | 82.65 | 79.49 | 78.45 | 85.54 | 13.27 |
| Ours | Base | 91.83 | 91.47 | 91.38 | 91.27 | 91.21 | 91.20 | 90.92 | 90.50 | 90.12 | 91.10 | 1.71 |
|  | Novel | - | **89.14** | **86.63** | **78.55** | **72.07** | **70.41** | **68.39** | **64.27** | **62.93** | **74.05** | 26.21 |
|  | Both | 91.83 | **91.29** | **90.70** | **88.73** | **86.42** | **85.09** | **83.41** | **80.84** | **79.24** | **86.39** | 12.59 |

As shown in Tables VIII to X, our method obtains AA scores of 37.91%, 93.31%, and 86.39% for both base and novel audio classes (the rows of Both) on the FSC-89, NSynth-100, and LS-100, respectively. These AA scores are higher than the counterparts achieved by the baseline methods. Our method produces PD scores of 8.21%, 12.90%, and 12.59% for both base and novel audio classes (the rows of Both) on the FSC-89, NSynth-100, and LS-100, respectively. These PD scores are lower than the counterparts obtained by the baseline methods. That is, our method outperforms all baseline methods in terms of both AA and PD when evaluated on three audio datasets. The advantage of our method over the baseline methods in the two metrics above mainly benefits from the STDU in base session, and the design of APGM and PQAM for updating prototypes. These three modules above work together to effectively reduce the confusions between the prototypes of various audio classes. Compared with the baseline methods, the proposed method can recognize novel audio classes better and forget old audio classes less.

In addition, the AA scores obtained by different methods on the FSC-89 are lower than that on the NSynth-100 and LS-100 for all methods in all sessions. The reasons are probably that the background noise in the FSC-89 is much stronger than that in other two datasets, and the sources of samples of the FSC-89 are less consistent compared to other two datasets. Hence, the inter-class confusion and intra-class inconsistency of the FSC-89 are larger than that of the NSynth-100 and LS-100.

To observe the confusions among different audio classes in the last incremental session, we plot the confusion matrices obtained by different methods on the LS-100, as illustrated in Fig. 6.

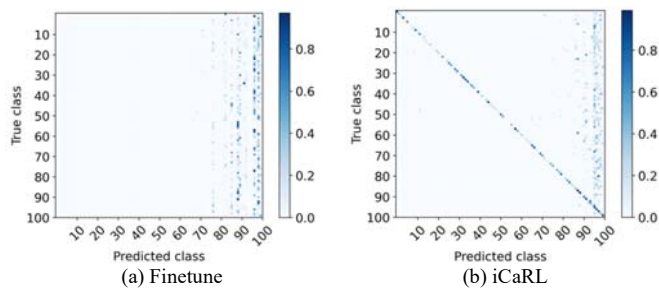

(a) Finetune  (b) iCaRL

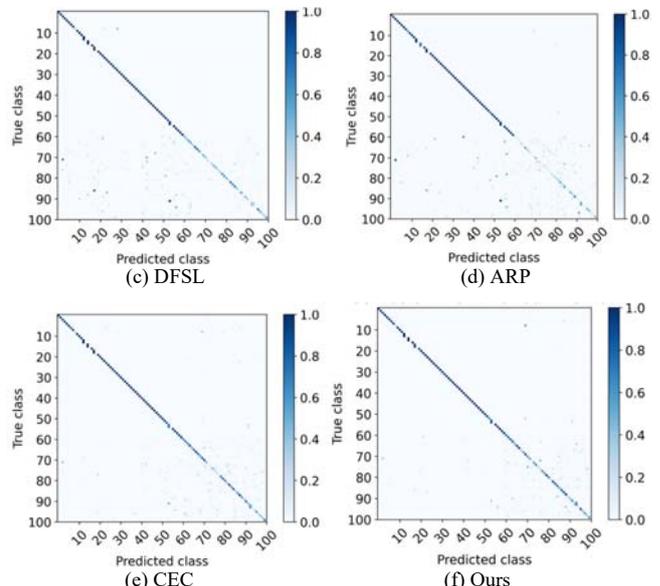

(c) DFSL  (d) ARP

(e) CEC  (f) Ours

Fig. 6. Confusion matrices of the last incremental session obtained by different methods on the LS-100.

In the confusion matrices obtained by different methods, most values lie in the diagonal which denotes the ground-truth, and the confusions among base audio classes (classes 0 to 59) are obviously less than that among novel audio classes (classes 60 to 99). In addition, our method generates a less scattered and lighter confusion matrix, which shows our method obtains higher accuracy scores (less confusions) than other methods.

### E. Computational Complexity and Memory Requirement

Because the training of initial classification system in the base session is generally conducted on a computing machine with high performance, the training time of the initial classification system is not critical for the problem of FCAC. However, the computational efficiency of different methods in incremental sessions is important in practice. Hence, the training time of different methods in incremental sessions on the LS-100 is recorded for computing their ATT values. For fair comparison, only the training time of different classification systems is included, whereas the time for data preparation and embedding learning is excluded.

This experiment is also conducted on the computing machine introduced in section III.*B* but only one GPU of RTX3090 is adopted here. It can be known from the second column of Table XI that the ATT of the Finetune method, iCaRL method, DSFL method, ARP method, CEC method and our method is 29.39 s, 144.20 s, 0.16 s, 0.48 s, 0.93 s and 0.84 s, respectively. That is, the computational complexity of the Finetune and iCaRL methods is much higher than that of other four methods. The reason is that the Finetune and iCaRL methods need many epochs to fine-tune the entire classification system in order to obtain satisfactory results. In converse, other four methods only need one or two epochs to update the classifiers (or the prototypes) instead of the entire classification system. In terms of ATT, our method has advantage over the baseline methods except the DFSL and ARP methods.

In practical application, samples may involve privacy and the storage space of intelligent audio terminals is usually limited. Hence, it is often not allowed to store abundant samples (or embeddings) and too many parameters, such as prototypes, weights. We compare the memory requirements of different methods in incremental sessions only, because the memory requirements of different methods in base session have been determined after the initialization of the classification system. The memory requirements (i.e., SS) of different methods in incremental sessions are listed in the third column of Table XI.

Because the Finetune method only needs to tune the parameters of the classification system obtained in last session using samples of novel audio classes without storing any samples and parameters for novel audio classes, the SS of the Finetune method in incremental sessions is always equal to 0. The iCaRL method needs to store some representative samples for each novel audio class in all incremental sessions. The SS of the iCaRL method is equal to $N_c \cdot K_s \cdot L_s$, where $N_c$, $K_s$ and $L_s$ denote total number of audio classes in all incremental sessions, the number of samples per audio class and the length of one sample, respectively. Our method, the DFSL method, the ARP method and the CEC method need to store one prototype (or weight vector) for each novel audio class in all incremental sessions. The SS of these four methods is equal to $N_c \cdot D$, where $D$ denotes the dimension of one prototype (or weight vector). $D$ is generally smaller than $L_s$, and thus the memory requirements of these four methods are lower than that of the iCaRL method in incremental sessions.

TABLE XI
AVERAGE TRAINING TIME AND STORAGE SPACE OF DIFFERENT METHODS IN INCREMENTAL SESSIONS

| Methods | ATT (s) | SS |
|---|---|---|
| Finetune | 29.39 | 0 |
| iCaRL | 144.20 | $N_c \cdot K_s \cdot L_s$ |
| DFSL | 0.16 | $N_c \cdot D$ |
| ARP | 0.48 | $N_c \cdot D$ |
| CEC | 0.93 | $N_c \cdot D$ |
| Ours | 0.84 | $N_c \cdot D$ |

*F. Generalization across Datasets*

In all experiments above, the training data and the evaluation data are selected from the same audio dataset. To evaluate the generalization capability of our method across audio datasets, the training data and the evaluation data come from different audio datasets. That is, when the training data is chosen from one audio dataset (e.g., FSC-89), the evaluation data is selected from the remaining two audio datasets (e.g., NSynth-100 and LS-100). In this experiment, the value of (*N*, *K*) is also set to (5, 5) without losing generality.

In the first row of Table XII, the item on the left side of the arrow (e.g., "FS" in "FS→NS") represents the training data, while the item on the right side of the arrow (e.g., "NS" in "FS→NS") denotes the evaluation data. The AA scores of all sessions obtained by our method across audio datasets are presented in Table XII. Our method obtains AA scores of 40.31%, 39.37%, 83.83%, 77.50%, 78.71%, and 77.31% for the class of Both, when audio datasets are FS→NS, FS→LS, NS→FS, NS→LS, LS→FS, and LS→NS, respectively.

TABLE XII
AVERAGE ACCURACY SCORES OF ALL SESSIONS OBTAINED BY OUR METHOD ACROSS AUDIO DATASETS (IN %)

| Class | FS→NS | FS→LS | NS→FS | NS→LS | LS→FS | LS→NS |
|---|---|---|---|---|---|---|
| Base | 41.45 | 41.51 | 99.90 | 99.95 | 91.87 | 91.45 |
| Novel | 38.89 | 37.73 | 21.53 | 10.09 | 23.68 | 38.18 |
| Both | 40.31 | 39.37 | 83.83 | 77.50 | 78.71 | 77.31 |

FS: FSC-89; NS: NSynth-100; LS: LS-100.

As given in Tables VIII, IX, and X, our method obtains AA scores of 37.91%, 93.31%, and 86.39% for the class of Both, when audio datasets are FS→FS, NS→NS, and LS→LS (training and evaluation data from the same audio datasets), respectively. The AA score of 37.91% (FS→FS) is lower than the AA scores of 40.31% (FS→NS) and 39.37% (FS→LS). However, the AA score of 93.31% (NS→NS) is higher than the AA scores of 83.83% (NS→FS) and 77.50% (NS→LS). Similarly, the AA score of 86.39% (LS→LS) is higher than the AA scores of 78.71% (LS→FS) and 77.31% (LS→NS). That is, our method achieves better results when the training data (except the FSC-89) and the evaluation data are from the same audio datasets. Furthermore, when the training data is from the FSC-89, even if the training data and the evaluation data are from different audio datasets, our method produces larger AA scores. The reasons are probably that samples in the FSC-89 are relatively noisy (with evident background noise) and sources of samples in the FSC-89 are diverse. Hence, the distribution range of time-frequency characteristics of samples in the FSC-89 is wider and may overlap with that of samples in the LS-100 and NSynth-100. Therefore, the classification system which is trained on the FSC-89 performs better on the LS-100 and NSynth-100. In summary, our method generalizes well across audio datasets instead of overfitting on a single dataset.

*G. Extended Analyses*

The first extended analysis is to discuss the settings of *N*-way *K*-shot for training the PAN in base session and for updating the classifier in incremental sessions. Specifically, we fix the number of query samples as 15 ($K_q$=15) and all modules of STDU, APGM and PQAM are adopted in this experiment. We discuss the impacts of the values of both *N* and *K* on the performance of our method. We set the same values of $K_q$, *N* and *K* for training the PAN and updating the classifier. We select the values of both *N* and *K* from {1, 5, 10, 15, 20}. The accuracy of the last incremental session obtained by our method on the LS-100 is presented in Fig. 7. The following observations can be obtained from Fig. 7. First, when the value of (*N*, *K*) is equal to (5, 5), our method obtains the highest accuracy score of 82.50%. Second, for the same number of ways, the larger the number of shots, the higher the accuracy scores (except 5-way 5-shot). The reason is probably that with the increase of shots, the classification system obtains more information about novel audio classes and thus obtains higher accuracy scores. Third, for the same number of shots, when the number of ways is equal to 5, our method obtains the highest accuracy score (except 20-way 20-shot). When the number of ways deviates from 5, the accuracy scores obtained by our method decrease. The possible reasons are as follows. When the number of ways decreases, the number of incremental sessions will increase and thus the old audio classes are more likely to be forgotten (the more sessions, the more likely the old audio classes will be forgotten). When the number of ways increases, the number of novel audio classes in one incremental session will increase and thus the confusions between novel audio classes in each incremental session is more likely to increase (the more audio classes, the greater the possibility of confusion between audio classes).

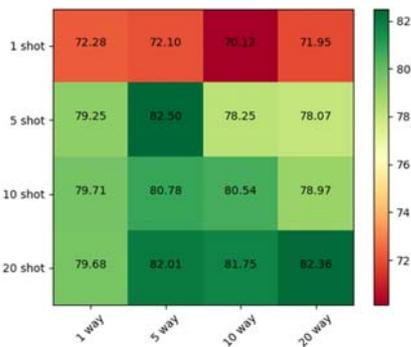

Fig. 7. Accuracy of the last incremental session obtained by our method on the LS-100.

The second extended analysis is to visually demonstrate the locations of query embeddings and prototypes before and after being updated by the PAN. We use the t-SNE [61] to map query embeddings and prototypes into two-dimensional space as depicted in Fig. 8. The Python library of *scikit-learn* is used to reduce the dimensionality of query embeddings and prototypes. The Python library of *matplotlib* is adopted to plot Fig. 8. Five audio classes are randomly chosen from the LS-100 as the base audio classes, and five new audio classes are added as the novel audio classes in incremental session. It can be observed from Fig. 8 that prototypes of the classifier are shifted away from the confusion region by the PAN to produce more discriminative decision boundaries when novel audio classes are involved.

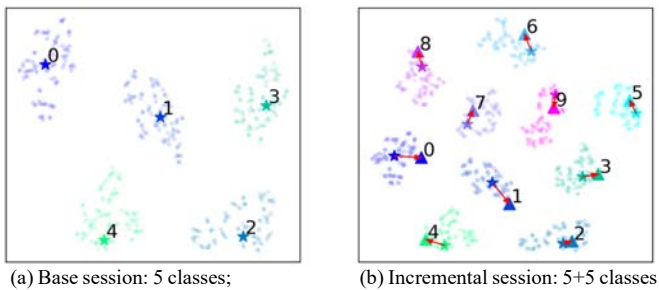

(a) Base session: 5 classes;  (b) Incremental session: 5+5 classes

Fig. 8. Visualization of query embeddings and prototypes before and after being updated by the PAN. Dots of different light colors denote query embeddings of various classes. The asterisks and triangles represent prototypes before and after update, respectively. Red arrows indicate the movement of prototypes caused by the PAN.

## IV. CONCLUSIONS

In this study, we have investigated a newly-emerging problem of FCAC. Moreover, we have tried to solve this problem by designing a dynamically expanded classifier with self-attention modified prototypes. Based on the detailed description of our method and comprehensive experiments and discussions, the following two conclusions can be drawn.

First, our method exceeds previous methods for FCAC in terms AA and PD under the same experimental conditions. As a result, our method is a state-of-the-art method for solving the problem of FCAC. In addition, our method has advantage over most baseline methods in terms of memory requirement and computational complexity.

Second, we design a PAN for updating prototypes of the classifier in incremental sessions. The PAN is a self-attention network and can effectively take advantage of prototypes of prior sessions and unlabeled query samples of current session for updating all prototypes of the classifier. In addition, we propose a STDU in base session to train the EE and PAN, which makes the EE and PAN possess better generalization capability in incremental sessions.

Although our method has advantages over baseline methods, there is still room for improvement in this work. For example, we did not update the EE in each incremental session and thus the generalization capability of the EE needs to be enhanced. In addition, we did not consider the implementation of the proposed method on intelligent audio terminals with limited computing resources. The future work will include two parts. First, we will design a strategy to update the EE together with the classifier in incremental sessions for further improving the performance of our method. Second, to meet requirements for lightweight applications, we will decrease the computational complexity and memory requirement of our method by taking effective measures, such as embedding grouping, network quantization, self-knowledge distillation. Accordingly, we can make the proposed framework lighter for directly deploying our method on intelligent audio terminals.